# Vortex ring generation during drop impact into a shallow pool

Andreas Wilkens[1], David Auerbach[2], GertJan van Heijst[2]

**Abstract**

When a drop falls from a moderate height into a shallow fluid pool whose depth is of the order of the drop-radius two pairs of vortex rings are generated. The inner pair forms at the edge of the crater created by the impacting drop while the outer pair is laid off from the spreading wave. One ring of each of these pairs is short-lived while the other persists. Each of the rings has a measureable non-zero circulation (whose sign, however, is opposite to that of the well-known deep-water drop vortex rings) which persists long after the wave has receded. Furthermore under certain conditions they develop instabilities typical to ring vortices. Although the rings are well reproducible, aspects of their later development are very sensitive to changes in some of the experimental parameters. This paper reports on experiments distinguishing reproducible aspects of these drop- generated rings and documenting their dependence on material and geometrical parameters.

## 1. Introduction

Ring vortex motion during drop impact has been the subject of extensive investigation, its forms depending on drop and pool substance, on the drop diameter and release height, and on the depth of the pool, the latter varying from a thin film (or a dry surface) to a pool whose depth may be many drop radii deep. Both these extremes have been studied rather extensively, and details of the crater (Berberović, 2009), the splash (crowning, jetting, waves, see *e.g.* Vander Wal *et al.* 2006) as well as the flow initiated beneath the surface (ring vortex, secondary vortex rings, see *e.g.* Saylor & Grizzard, 2006) have been documented. Mohamed-Kassim & Longmire (2003) performed experiments on drop impact into a pool, concentrating on the deformation and the vorticity distribution during the first moments of impact. Okawa *et al.* (2006) studied drop impact into a shallow pool, but mainly examined aspects of the splash. Manzello & Yang (2002) examined this aspect too, concentrating however on the effect of using different pool fluids. Tan & Thoroddsen (2002) let drops fall into glycerol and investigated Marangoni instabilities when the flow is initiated by the surface tension gradient between water and glycerine.

A feature common to drop impact studies, whether they be into shallow or deep pools, is that the phenomena can be divided into aspects which are well repeatable and those which are unique for each drop. In our case although the existence, initial form and the motion of these rings are repeatable, the evolution of possible instabilities and subsequent interaction and metamorphosis are particularly sensitive to minute changes in the initial conditions of drop impact as well as to material properties near the pool surface (e.g., surface tension and viscosity, film pressure).

Now for repeatable aspects drop impact may be characterised by the following physical parameters: $V_0$, the impact velocity (or, alternatively, the drop release height *h*), the drop diameter *D*, the pool depth $h_d$, the gravitational acceleration *g,* and the density *ρ*, the viscosity *µ* and the surface tension *σ* of the liquid. These seven parameters may be expressed in terms of three different dimensions (mass, length, time), so that according to the Buckingham Theorem the problem is characterised by a total of four independent non-dimensional numbers. A common choice is the following: the Reynolds number $Re = \rho V_0 D/\mu$, representing the ratio of inertia and viscous forces; the Froude number $Fr = V_0^2/(gd)$, representing the ratio of inertia and gravity; the Weber number $We = \rho d V_0^2/\sigma$, denoting the relative importance of inertia and surface tension; the geometrical aspect ratio $H = h_d/D$, being the ratio of the pool depth and the drop diameter. In some studies one uses the Ohnesorge number $Oh$, which is defined by a combination of $We$ and $Re$, namely $Oh = \sqrt{We}/Re$, which relates the viscous forces to inertial and surface tension forces. These scaling factors have proved useful in reducing the number of parameters necessary to categorize impact features (e.g., splash dynamics).

---

[1] Institut für Strömungswissenschaften, Herrischried/Germany
[2] Fluid Dynamics Laboratory, Department of Physics, Eindhoven University of Technology, The Netherlands

Here we will concentrate on reproducible flow structures generated at and beneath the surface for the case of drops with a moderate impact velocity and pools with a depth of the order of the drop radius. Water was used as drop fluid and the pool fluid was either water or glycerol.

## 2. Experimental setup

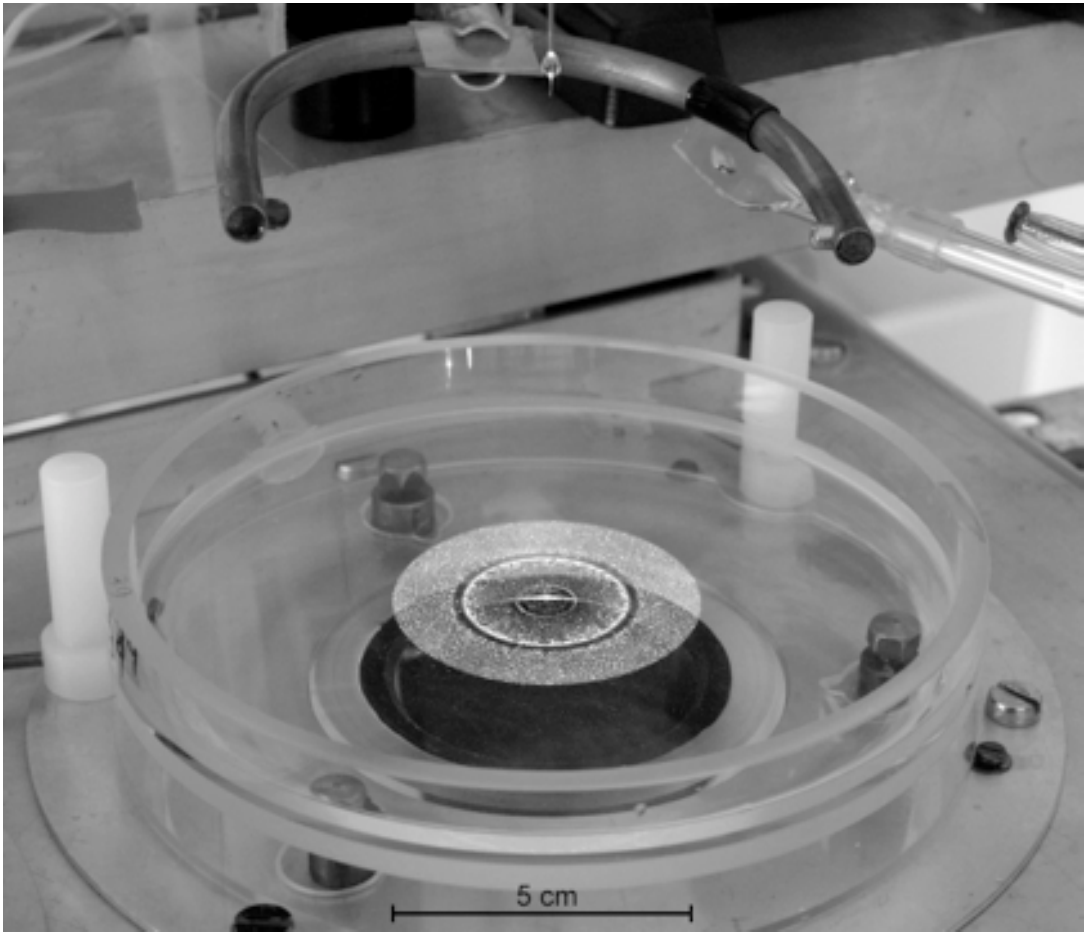

Figure 1. Experimental setup showing the Petri dish with distilled water (not visible). Chalk particles are visible on the bottom of the dish. The photoelectric switch is visible above

The drop apparatus consisted of a specially built glass Petri dish of 14 cm. The upper and lower surfaces of the glass bottom were carefully checked to be parallel (and to the horizon – i.e. horizontal - too, so that they are also parallel to the upper water surface), one of the important prerequisites for the Toepler-Schlieren system used here (see, e.g., Liu et al, 2008). The dish contained a pool of water into which the drops were released, as shown in Figure 1. The drop release mechanism consisted of an injection needle whose one end was carefully machined (visible at the top) and whose other end was connected to a reservoir via a valve. A photoelectric switch is visible at the top of the photograph shown in Figure 1. The switch was connected to a computer, which controlled the delay for the flash to an accuracy of better than 0.1 ms. Single images were thus taken and, for the case of series, each series represents different drops taken at different times. This method obviously demands a high degree of repeatability. This was indeed the case for the wave motion as well as the aspects of the ring structures that we study here. We shall discuss this issue in more detail presently.

Four different optical setups were employed to render various aspects of the phenomena visible: *light reflection* on the water surface waves, *seeding particles* (chalk, lycopodium), *dye* (ink) and *Schlieren* optics. Comparing results across visualization methods helped in gaining a fuller picture.

For the first method light was reflected at an angle of 30° to the water surface to render details of the waves visible. A variation, which often helped, used shadowgraphy with an asymmetric light source. This method has the advantage of not interfering with the flow, and may thus be used for determining structures resulting from drop impact of pure water drops into a pure water pool. It yields information on the motion and pressure field. When used alone the information from this method is sparse, so that it requires information from other methods for its interpretation. For a discussion of advantages and disadvantages of this method, see *e.g.* Willert & Gharib (1997) or Brocchini & Peregrine (2001).

For the second method seeding with either washed lycopodium powder or washed chalk powder gave details of the motion beneath the surface and near the bottom in particular, depending on how long the chalk was allowed to sink. Dark-field optics was used for obtaining particle paths (with a 0.1 mm light sheet from below viewed obliquely at 45° for the floor and at 40° for the volume). An advantage of lycopodium powder was its large albedo compared to that of chalk. However, even after washing the lycopodium powder released substances, which affected the surface.

For the dye method we used Pelican ink in a 1/20 dilution, using either dyed drops and a clear pool or *vice versa*. The light source was diffuse and asymmetrically placed in order to obtain additional information on the shape of the surface (wave, crater). Although aspects of motion and mixing can be determined relatively well in this way, one of the main problems for our setup was the extent to which (probably the surface viscosity of) the ink changed the flow pattern. A combination of reducing the ink concentration and employing other visualization methods allowed us to estimate the extent of this essentially surface viscosity effect.

For the Schlieren optical method distilled water drops fell into a pool containing water with 13% glycerine. Since Schlieren visualization is based on refractive index gradient, this method worked best after several drops had fallen into the pool and was most useful for visualizing mixing aspects of the flow. This method leads to a change in material parameters from drop to drop and has a significant effect on details of the flow in the core, various instabilities and the flow near the edge of the vessel, none of which we discuss here. The motions, which we discuss, are hardly affected by the additional drops and we discuss the case of pure water too in detail.

In fact, since only the light-reflection method for visualizing the surface waves is really interference-free one might well ask the question as to how relevant these visualizations are to the case of pure water drops falling into pure water. As mentioned in the introduction, reconstructing flow structures on the basis of how their pressure field is indicated by surface signatures is fraught with problems, as discussed, for example, by Willert & Gharib (1997) and Brocchini & Peregrine (2001). However, we learnt how the rings visualized via the other methods appeared in the wave images and then found the same wave images in absence of any foreign substances. Secondly, dilution (less glycerine, powder, dye) did not alter essential features of the structures as long as the contrast was visible. Finally, each of the methods interferes with the flow in its own characteristic manner, yet all indicate the existence of the ring structures as will be described in the next section.

## 3. Results

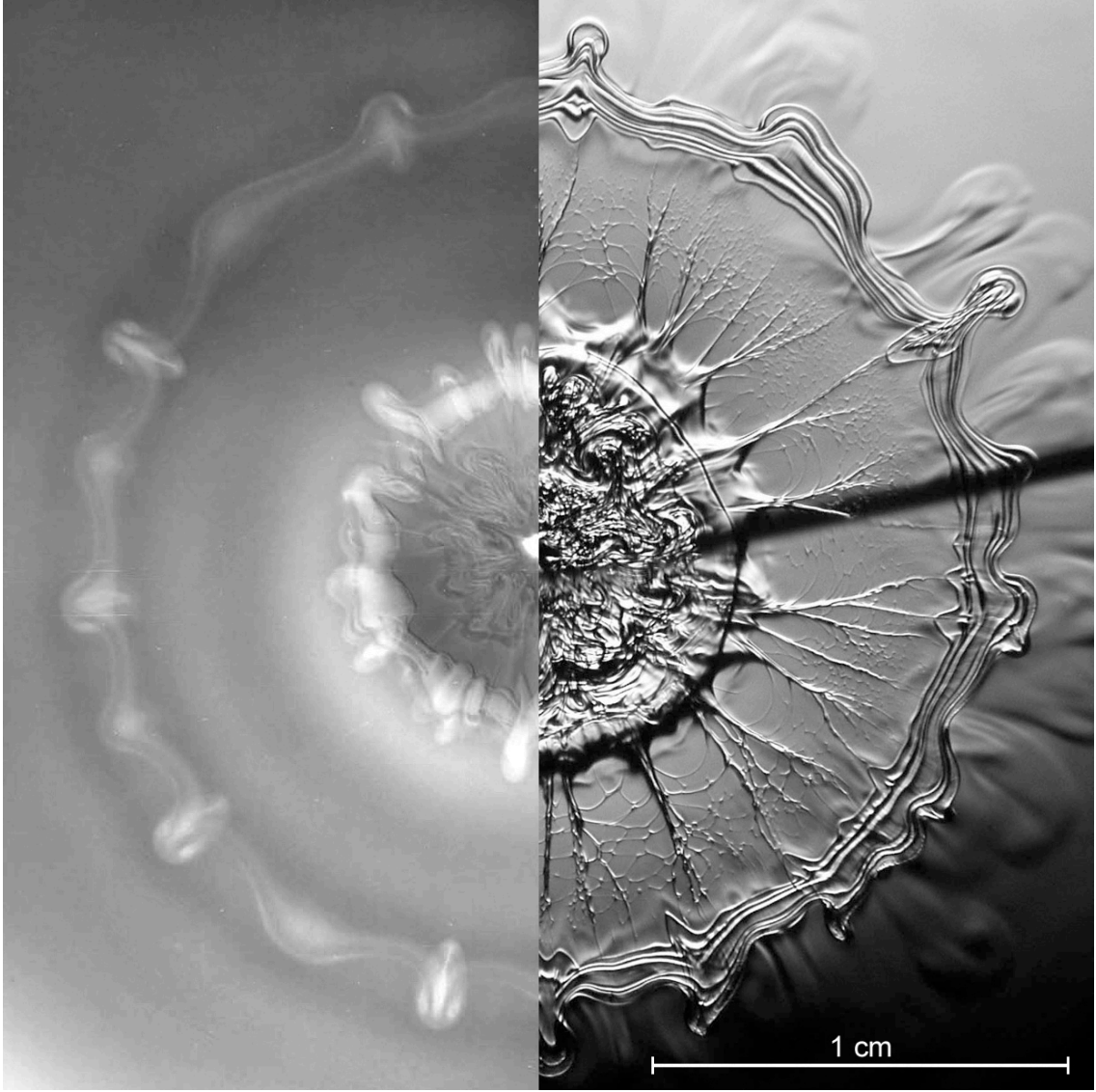

Figure 2. Composite image taken from above 100 ms after impact. Left: dye visualization of the flow. Right: Schlieren image after 9 prior drops had fallen. The dark line on the right is the shadow of the injection needle.

Figure 2 shows visualizations of a flow pattern caused by pure water drops with diameter $D = 3$ mm falling from a height of 97 mm (drop underside to pool surface at the centre) into a pool of water-glycerine (13%) with depth $h_d = 1.1$ mm, and impinging onto the free surface at a speed of approximately $V_0 = 1.4$ m/s. This corresponds to $Fr = 63$, $Re = 4367$, $We = 82$, and $H = 0.37$ (and $Oh = 0.002$), indicating the relative importance of inertia, gravity and surface tension. Most of the experiments reported here were carried out for these experimental parameter values and if not specifically mentioned, these values apply to all experiments discussed below. The composite image presented in Figure 2, showing visualization with a dyed pool (left) and with Schlieren optics (right), was taken from above at $t = 100$ ms after impact (taking place at $t = 0$ ms). The image on the right is a Schlieren image taken after 9 prior drops had fallen into the pool (determined via trial-and-error as yielding an optimal contrast between pure water and glycerol). The two images match well, both sides showing an inner more chaotic region and an outer ring with nodelike structures. These structures are initially insignificant and generate through an impinging axial flow from either side of the nodes, causing them to enlarge and deform. The dye-image thus renders the unstable ringlike structures at the edge of the inner region more clearly, whereas the Schlieren image better reveals the mixed inner core, the more or less radially emanating dendritic structures between the two rings and regions of slightly varying density outside of the outer ring. The inner ring generally contained more drop fluid while the outer contained more pool fluid. The dendritic structures and the nodelike structures (number, form, position, and development) are the least reproducible phenomena, changing significantly from drop to drop. Their dynamics is the subject of on-going work. We focus now on a cross section of the outer ring, aspects of which are well repeatable.

Figure 3 shows particle paths illuminated by a vertical light sheet viewed at an angle of 40°. It is thus an oblique view of the vertical cross-section of the right-hand side of the pool, with the pool centre lying to the left of the image. The particle streaks in this image (taken at $t = 34$ ms;

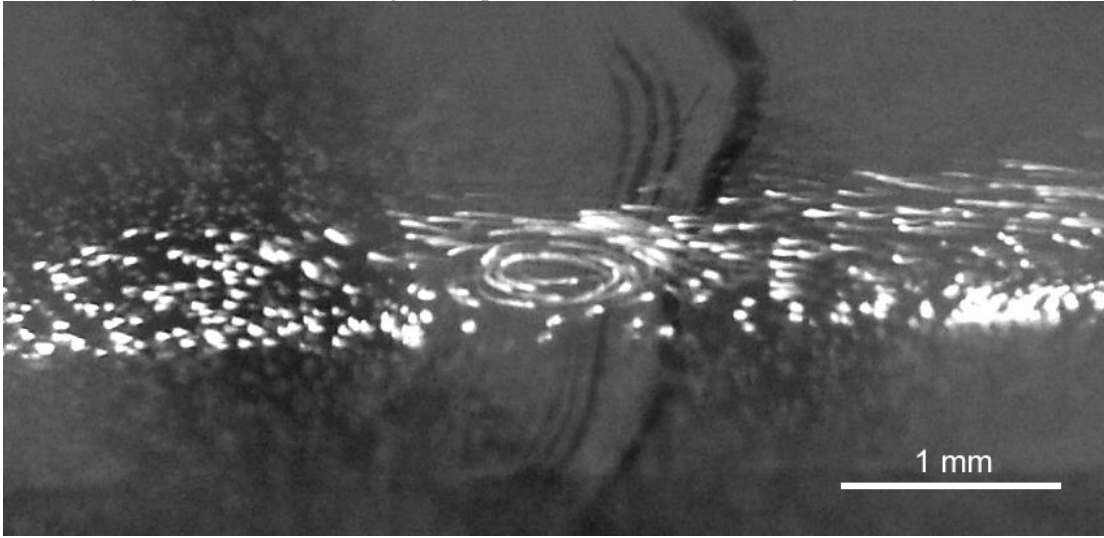

Figure 3. Particle paths illuminated by a vertical light sheet viewed at an angle of 40° showing the outer vortex clockwise motion (exposure time: 3.3 ms). The pool centre lies to the left of the figure, floor and pool surface corresponding approximately to the upper and lower extent of the tracers. The superposed Schlieren image (exposure time: 0.01 ms, hardly visible) taken obliquely shows the core (slightly displaced due to the perspective), indicating the outer vortex instability

exposure time: 3.3 ms) indicate that there is a local turning motion associated with the outer ring. Since the brightest part of each streak corresponds to the beginning of the path, there is an apparently negative circulation with which the ring may be associated (in clockwise direction – in the following we consider the right side only as our circulation sign convention). We have superposed the corresponding Schlieren image (exposure time: 0.01 ms) taken obliquely, which shows an azimuthally wavy structure. When the particle path image is corrected for obliquity and refraction (significant when the wave is present above the ring. A Zeiss Abbe-type refractometer was used to estimate the mean change) the apparently elliptical ring is approximately circular, having its axis at half the pool depth. This centre is some 10 mm distant from the impact centre. The inner ring was observed to appear some 10 ms after impact while the outer ring first appeared after 14 ms. How large are their circulations? As can be seen from the photograph, the fastest-turning particles (corrected value of about 0.2 mm from the centre) trace an arc of some 90° per 3.3 ms exposure. This represents a circulation of around -100 mm$^2$/ s. However, not only do other problems of 3-dimensionality (e.g., waviness, radius determination) but also the lightening and darkening of the flash render an error in the circulation of around 50% (see discussion later).
A similar inner ring appears at the edge of the crater (visible in Figure 2, but not in Figure 3) at a distance of about 4.5 mm from the impact centre. Analysis of streaks showed that the circulation senses of the inner and outer rings are the same, and also that the magnitudes of their circulation are comparable (see Figure 7 later for a sketch of these structures). As mentioned the wave (see later Figure 8) has since left the ring region, being some 35 mm from the centre.

The inner ring arises at the lower edge of the crater created by the drop impact. This occurs when the crater begins to collapse (around 10 ms after drop impact for this configuration). By this time the expanding wave has moved some 6.5 mm from the centre, as shown in Figure 4.

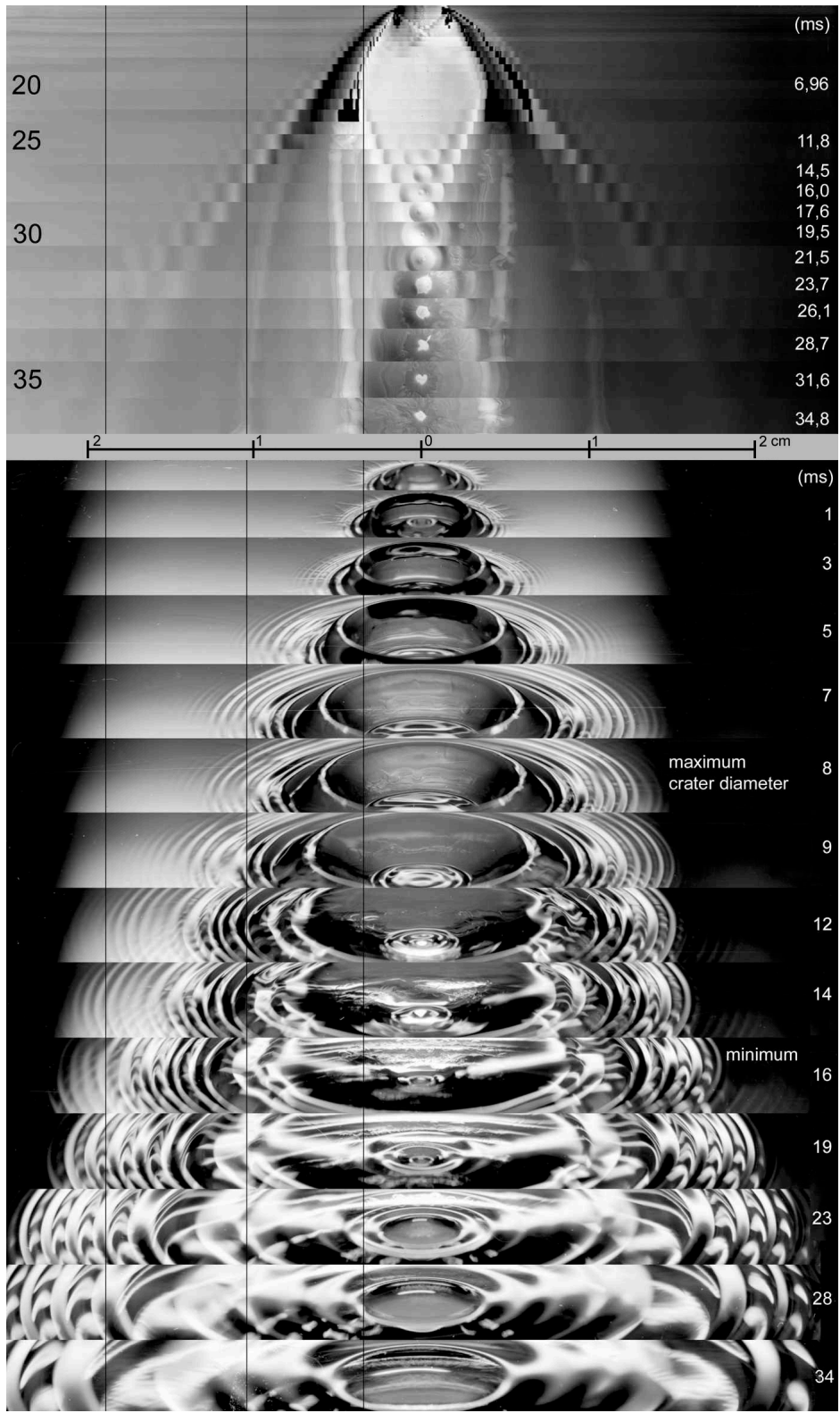

Figure 4 shows the flow due to undyed water drops impinging serially into a pool of (above) slightly (5%) dyed 13% glycerol and blow, 13% glycerol alone.

Figure 4 shows the flow due to undyed water drops impinging serially into a pool of (above) slightly (5%) dyed 13% glycerol and blow, 13% glycerol alone (the use of water instead of glycerol gives essentially the same images, albeit far less clear). This figure is a composite image of surface waves recordings (lower part) combined with a sequence of dye visualization recordings made with a dyed pool (upper part). The lower series was taken from an angle of 70°, the upper, from directly below. The image number is shown on the left while the numbers on the right give the time in ms after drop contact. The numbers on the right give the time in ms after drop contact (image number on the left). A scale (in cm) for the radial extent divides the two composites. The lower image shows activity on or near the surface: the creation and collapse of the crater and the motion of the expanding wave. The upper image is a composite of horizontal strips of such images as the left side of Figure 2 for increasing times, their height corresponding to the time interval between images. They thus show both the positions of the rings and those of the waves (from which velocities can be deduced). The quickest (outer) expanding bright region is the main wave.

The innermost dark stripe (dark on the left, bright on the right due to asymmetric lighting) on the other hand, depicts the front of the refilling crater during collapse. The point at which it crosses (image 30 left) marks a further wave which moves from the centre radially outwards as the crater level at the centre exceeds its original height. Both the inner and the outer rings appear as the main bright stripes between the waves. It is observed how the inner ring, consisting largely of drop-fluid, grows and then gradually shrinks (ever bounding the thoroughly mixed inner region). The outer ring, consisting largely of pool fluid, expands at an ever-diminishing rate. The waves move radially outwards with a maximum velocity of around 23 cm/s at the onset, slowing down and flattening out as they move. After about 0.44 s they again reach the centre after reflection at the circumference.

It is interesting to note that the wave structure alone indicates not only the rings, but also their instabilities (these are far clearer in the Schlieren and dye images discussed earlier).

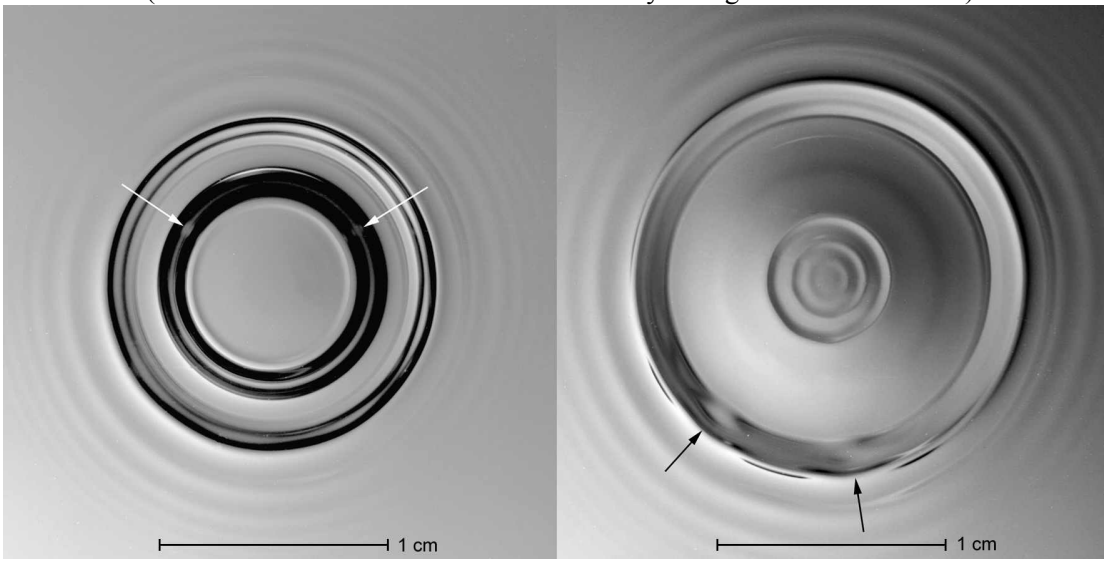

Figure 5. Expanding waves showing the onset of instability. The arrows indicate the location of the instabilities on the vortices (a) on the (primary) inner vortex at t = 9.6 ms, and (b) on the (primary) outer vortex at t = 11.8 ms after impact of a pure-water drop into pure water.

Figure 5 gives an example of the onset of instability for the case of a pure water drop falling into a pure water pool. The arrows indicate the location of the instabilities on the rings (a) on the inner ring at $t = 9.6$ ms, and (b) on the outer ring at $t = 11.8$ ms after impact.

Using the dye method one notices that the two above (primary) rings spawn weaker secondary rings. The two pairs of white rectangles for each part of Figure 6 indicate each primary ring with its corresponding secondary ring, long marks indicating primary, short ones secondary rings. We determined circulation of the secondary rings to have the opposite sign to that of the primary rings. The secondary rings were best rendered visible by the transport of newly dyed crater fluid (created by the fall of a new ink drop). Both secondary rings are short-lived compared to the lifespan of the primary rings. The outer ring first really becomes apparent around $t = 15$ ms and we estimate that it persists until about $t = 35$ ms, whereas the inner ring becomes apparent around $t = 10$ ms and persists until $t = 20$ ms. All the above mentioned four rings are well-reproducible aspects of the drop impact and their generation process and positions are schematically depicted in Figure 7.

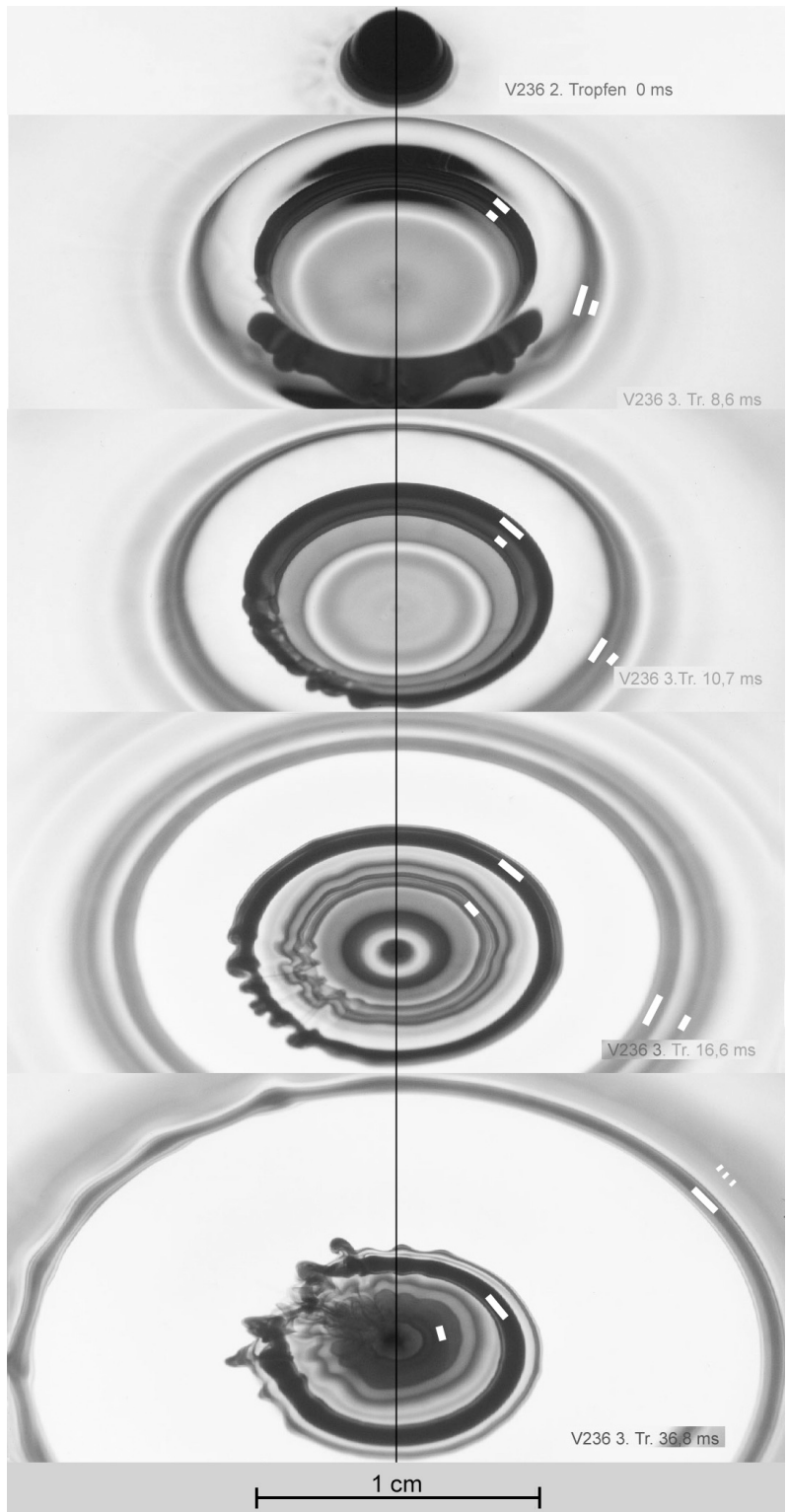

Figure 6. Dyed images showing both primary and secondary vortices. The numbers indicate elapsed time, the number of preceding drops (for optimal visualization), and the time after impact in ms. The white marks indicate the vortices: long marks for primary, short ones for secondary vortices.

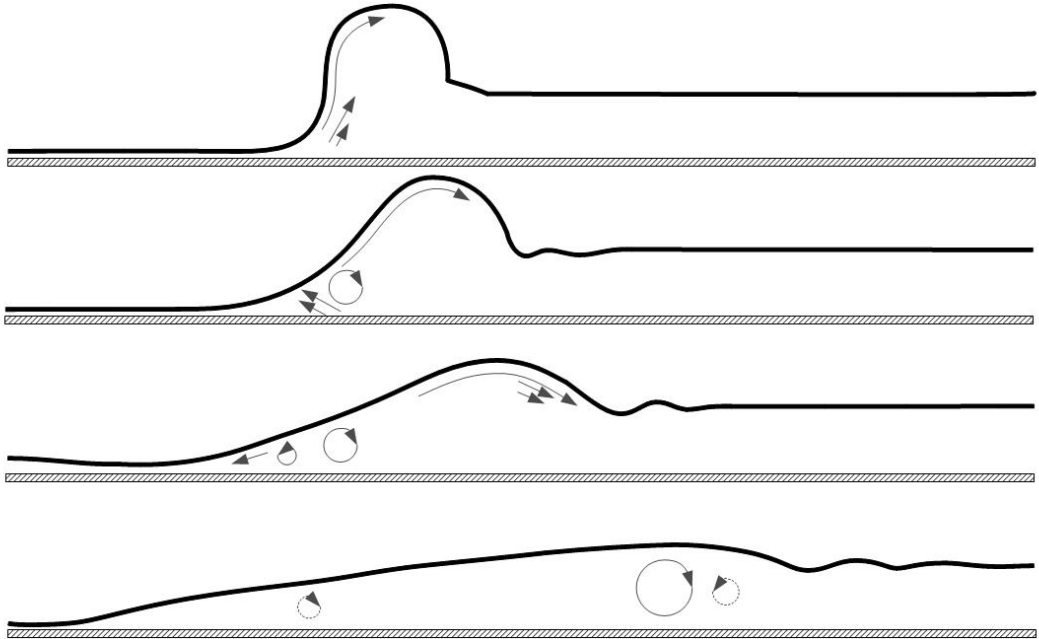

Figure 7. Diagram schematically showing radial cross-sections (only the right half is shown here) during the generation of the four vortex rings (a – d) 9, 12, 15 and 21 ms after drop impact. The left side marks the pool centre with the empty (7a) and the filling crater (7b-d), while the radially expanding wave, together with its capillary precursors, moves towards the right. For further details, see text.

Figure 7 shows a schematic diagram of salient features of the flow at four stages during the generation of the four rings (7a – d) 9, 12, 15 and 21 ms after drop impact. Radial cross-sections are depicted (only the right half is shown here), the left end marking the pool centre with the filling crater (crater wall in top figure 10mm from the centre). Figure 7a shows the expanding wave - together with its capillary precursors – moving towards the right. The arrows indicate flow directions gleaned from various images which we feel are pertinent to the phenomenon and which we shall discuss presently. Aspects of the dynamics become better explicable when one assumes them to be vortices (see later for a detailed discussion). Figure 7b shows the formation of the inner primary vortex as the wave begins to flatten. The two arrows left of inner vortex schematically show the flow induced near the wall by the inner vortex. The arrow near the free surface indicates a flow tangential to the surface. Figure 7c includes the inner secondary ring of opposite sign, possibly resulting from the flow induced in the sense of the two arrows in Figure 7b. The final stage, Figure 7d, shows the (decaying) inner primary vortex (the inner secondary vortex had by then cessated) and the outer primary and secondary vortices as the wave further flattens. The inner ring pretty much fills out the layer whereas the outer ring first becomes apparent towards the top of the layer in the wave, only filling out the entire layer when the wave has disappeared (*c.f.* Figure 3).

Figure 8, obtained from series of images like those collected in Figure 4, shows the measured radial motion of the crater front, the outer ring axis, the expanding wave, the inner ring axis and a wave crest structure, which is precursor to the outer ring. This figure allows us to trace the appearance, position and motion of the crater front and the expanding wave as well as the inner and outer rings.

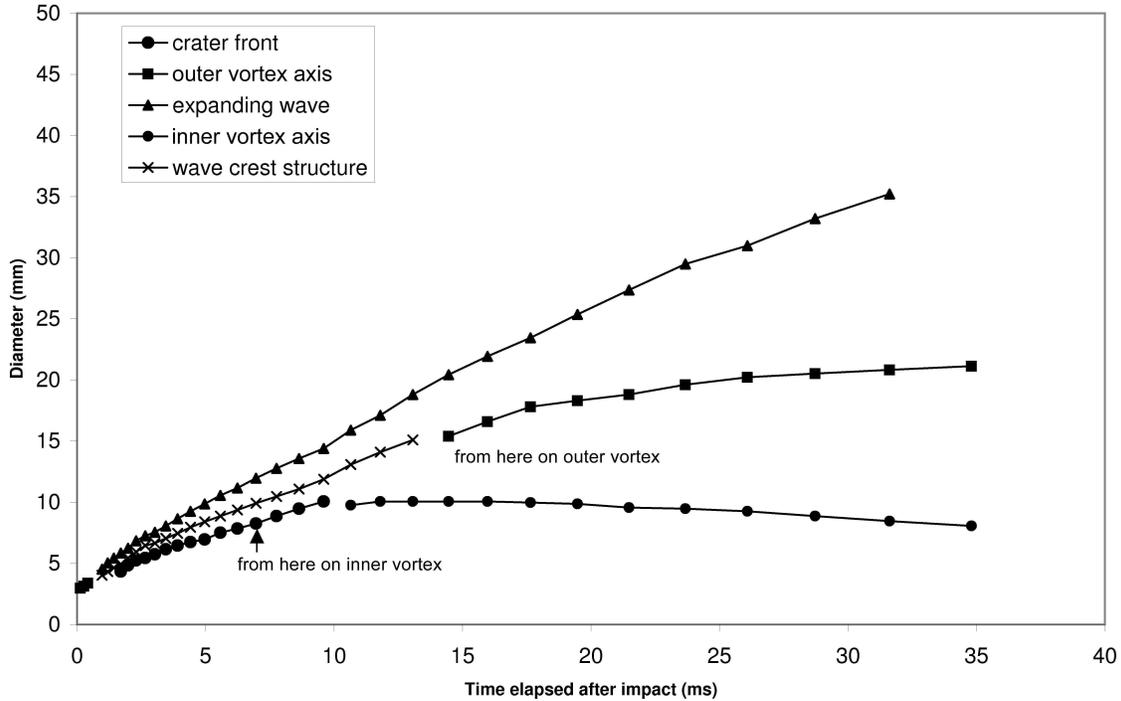

Figure 8. Time evolution of the position of waves and vortices during drop impact.

The sequence of events described above is well repeatable for the parameter combination mentioned. We now present results on the dependence of the outer ring diameter on various parameters. For this part of the experiment we used the seeding method with particles on the bottom. After drop impact, the ring sweeps away particles from the bottom as it moves along. This region is only slightly cleared after one drop and becomes ever more cleared as more drops fall, with the diameter of the cleared circular spot corresponding approximately to the final diameter reached by the outer ring, as indicated in Figure 8. A single image was thus taken after 40 drops had fallen and the entire wave and vortex flow had come to rest (see Figure 1).

Most of the experiments discussed up to here were performed for the same experimental parameter values, *i.e.* pool depth $h_d$ = 1.1 mm, drop Diameter $D$ = 3 mm, drop release height $h$ = 96 mm, and a 13% glycerine concentration of the pool fluid. In order to investigate the effect of changing these experimental parameters, additional experiments were carried out. In the following we present results on changing the pool depth, the release height (measured from lower drop surface to pool surface) and the drop diameter, as well as using a distilled water pool with no glycerine, and finally, on introducing surfactant. For all the following measurements care was taken to chose height/diameter configurations where the oscillation phase of the drop is as spherical as possible on the underside on drop impact (determined both by strobed shots as well as by video). We look at the effect on only one parameter, the outer vortex diameter.

The outer rings are clearly distinguishable and reproducible for pool depths $h_d$ between 0.65 mm and 1.43 mm (drop release height $h$ = 96 mm, drop diameter $D$ = 3 mm) and for drop release heights $h$ between 76 mm and 120 mm (pool depth $h_d$ = 1.1 mm, drop diameter $D$ = 3 mm). Increasing the pool depth between these bounds causes an increase in the outer vortex diameter from 25 mm to 29.3 mm (although these increases seem small, they represent the entire range for which the rings are generated). In the absence of glycerine (*i.e.*, pure water) the outer vortex responded most sensitively to a change in several parameters. Increasing the drop diameter $D$ from 2.9 mm to 3.3 mm (drop release height between 76 mm and 79 mm) caused an increase in the outer vortex diameter from 21.3 mm to 25.6 mm. Increasing the drop release height $h$ from 79 mm to 108 mm (drop diameter $D$ between 2.9 mm and 3.2 mm) caused the outer vortex diameter to increase from 21.3 mm to 27.3 mm. In the presence of glycerine the outer vortex diameter was generally somewhat smaller (5% - 8%) and in addition, became less sensitive to a change in the above parameters. Heating the water by 3K caused a similar decrease. The smallest drop diameter yielding an outer vortex was around 2.9 mm (drop release height $h$ = 79 mm), while drops with a diameter exceeding 3.3 mm (with $h$ = 96 mm) yielded a flow where the vortex structures were no longer repeatable.

The two most significant effects of introducing a surfactant (TPS) were the stabilization of the rings and the reduction in their diameters. By stabilization we mean that the instabilities mentioned earlier are supressed, so that they remained ringlike as they decayed. This stabilization may be due to a reduction in their intensity due to the viscous action of the surfactant. Another possibility is the changed interfacial conditions. A typical reduction in the diameter of the outer vortex was from 25.4 mm to 23.3 mm on introducing a 10 ppm concentration of surfactant. A further increase in the concentration of surfactant did not lead to a further decrease in the diameter. This 'saturation' effect is well known and seems to be due to the initial covering of the surface with a monolayer.

**4. Discussion**

*General considerations*

Distinguishing between stray vorticity and a concentrated vortex structure in an unsteady flow is a difficult and often more or less arbitrary business. This problem for a similar geometry has been discussed by Didden & Ho (1985) in the spirit of the Moore-Rott-Sears (MRS) criterion: unsteady separation occurs when both shear stress (and, of course) velocity vanish in a frame of reference moving with the separation point. Now if one looks at the motion of the inner and outer rings as shown in Figure 8 one sees that after 20 s at most they have both ceased to move, simply turning on the spot. For this situation the flow there has become steady, streamlines and pathlines coincide and the problems requiring the MRS criterion for the definition of a vortex are no longer necessary.
A second problem is that of distinguishing a vortex from the well-known circular Lagrangian motion of particles at and beneath waves. The idea is that some dye and other visualization artefact selectively render such Lagrangian flow beneath the wave visible. If this were the case, such Lagrangian-type circular motion would take on interesting new characteristics: Firstly, the ring motion does persist after the wave has receded. Secondly, waves grow on the ring and it takes on a form typical to well-known ring vortex instabilities (see, for example, Archer 2010). Thirdly, dye drains from the periphery of the ring into the core and an axial (azimuthal) flow along the ring's centre – so typical of a ring vortex instability – develops, one which (discussed above) drains fluid to the nodes. Finally, we mention the secondary structures discussed above, at least in part explainable by vortex interaction.
In the above discussion we have broached the broader problem of what a vortex is (see, e.g., Vétel et al., 2009 and Jeong & Hussain, 1995). The clearest evidence of these rings being vortices lies in the persistent more or less steady turning motion in one sense only, the appearance of an axis which becomes wavy and the draining of dye first from the ring periphery to the centre and then along the core typical to the classic ring vortex instability. However, details, such as the temporal evolution of the vortex strength, will need to be studied further.

*Scaling*

The vortex ring associated with drop impact into deep pools has the opposite sense to the primary vortices discussed here. The fact that this is a shallow pool has two important consequences. Firstly the motion of shallow waves is associated with a corresponding significant sloshing convective motion below the surface. Secondly for a shallow pool one will certainly expect the interaction of the flow with the floor to play a vital role in the ring dynamics. Now both pairs of rings were clearly distinguishable and reproducible for pool depths between 0.6 mm and 1.7 mm, becoming less significant or less distinguishable outside of these bounds. The diameter of the outer vortices increased both on increasing the pool depth as well as the drop radius. The question of scaling of the ring positions with the various experimental parameters thus still remains open.

*The core*

The core is the region into which the new drops fall. It is here that the impact takes place, here that the crater is initially formed and filled, and here that the most intense mixing of drop with pool fluid takes place. When a pure water drop falls into a pure water pool (Figure 5) none of the mixing, which takes place, is visible. The best impression of the intensity of the mixing is obtainable from Schlieren images (Figure 2). Due to the serial drops one can follow the mixing process, in particular, the extent to which drop fluid moves outward. We mentioned the fact that for the Schlieren method a dilution process takes place (not the case for the light reflecting method). Much work has been done on the effect of dilution on the various structures and it certainly has a marked effect as various dilutions are employed. The dilution due to the serial drops in the Schlieren method, however, occurs mainly in the core and the features of the rings, which we discuss here, are not significantly affected, as is evidenced by the pure water experiments.

*Vorticity source of the inner primary ring*

What is the source of the inner ring's vorticity? Now the sign of vorticity generated by the floor boundary layer of the radially outward flow is the same as that of the primary vortices (see short arrows in Figure 7a). This is thus a likely source of vorticity, prompting the question as to how this vorticity comes to detach from the floor. Two possibilities come to mind, either a countercurrent effect or the geometry of the crater being similar to that of a hydraulic jump. The idea of the first mechanism is that the fluid driven by the collapsing crater generates a countercurrent slumping flow, as described, for example by Das & Arakeri (1998). Such a flow is initiated as the crater crest begins to slump where the static pressure on the floor directly below the edge is highest. Such a flow might have the ability of detaching the boundary layer vorticity generated during the crater creation phase. The second possibility is that the outward flow just prior to the end of crater generation has a diverging character akin to the flow at a hydraulic jump. Such a flow is known to induce a vortex to detach from the boundary layer opposite the diverging boundary (*e.g.*, Liu & Lienhard, 1993).

A further possibility is that this detachment is of a Marangoni type as suggested by Tan & Thoroddsen (2002). They measured a circulation of around -5 mm/s$^2$ (at that radius – 0.2 mm, period, 0.3 s– *c.f.* their Figure 6) compared to our circulation of -100 mm$^2$/s (*c.f.* typical deep water drop vortex circulations at such times of +200 mm$^2$/s for drops laid directly onto the surface). Both our and their senses are negative. However, although Marangoni flow plays a part for different drop and pool fluids, rings are also generated for pure-water drops into pure-water pools too where surface tension gradients are absent. Marangoni induced flow cannot thus explain our inner vortex and we have to look elsewhere.

*Vorticity source of the outer primary ring*

How is the outer ring generated? As we have seen this vortex only becomes apparent when the wave has flattened significantly (Figure 7d). The small arrows near the surface in Figure 7c indicate the observed shear near the surface. An important question here seems to be as to what causes the vortex to be generated or 'laid off' in this highly repeatable manner. Several possibilities come to mind: breaking wave dynamics (*e.g.*, Duncan, 2001), criticality (hydraulic jump) or some form of ground effect. For traditional breaking waves vorticity is imparted the surface via wind shear and the vortex is concentrated on the crests of the waves. However, a comparison between the surface profiles of *bona fide* breaking waves and the waves in this case have little resemblance, for, after participating in the crowning, they flatten continuously as they move outward from the centre. It is also unclear which wind would be present, wind due to the approaching drop prior to impact - or some other surface tension or inertial effect. As to criticality the idea is similar to that used for the inner vortex: the laying off of the secondary vortex corresponds to a transition from a supercritical to a subcritical state in Froude's spirit, the former being able to transport vorticity, while the latter state forces the vorticity to be overtaken by the expanding wave. The third idea is that of the ground effect: Assuming the vorticity to be initially concentrated in the curve of the crest, *i.e.*, towards the upper part of the pool, it is then forced downward by the flattening wave. As it approaches the floor it will induce a flow in the sense of the method of images, *i.e.*, the floor will behave in the sense of an image vortex and this 'pair' will tend to move towards the centre, *i.e.*, in the opposite direction of the expanding wave. In conclusion although precursor structures are evidence that the outer vorticity has its origin during impact, this is an open question requiring further study. The fact that such vortex rings do not generate in deep water (above 40 mm) indicates the importance of more than one factor.

*Vorticity source of the secondary rings*

We have called the secondary vortex rings so because they are substantially weaker than the primary vortex rings. Their circulation sense is negative and is such that the primary inner vortex could have induced floor boundary layer fluid to detach whereas the primary outer vortex could have induced surface interfacial layer fluid to detach (see arrows in stage 2 of Figure 7, see, *e.g.*, Didden & Ho, 1985 or Walker *et al.*, 1987). A further study, however, will be necessary to decide this question. In particular the strength of the outer secondary vortex ought thus to depend on surface shear, something which we have not investigated, but which may well hardly exist.

*Non-repeatable aspects*

We have already discussed the core region and now draw the reader's attention to two further non-repeatable aspects of the drop impact flow, formation of the dendrites and the metamorphosis of the outer vortex. As to the former structures, which are best seen in the Schlieren images (right side of Figure 2), they emanate from the core and, with each successive drop, although changing their form, gradually settle down to take on an ever more constant form. The latter structures have already been discussed as being typical to a Crow-Widnall type of ring vortex instability. We used this name to draw attention to known similar forms during vortex impact. However, this flow is most probably more complex: Even the notion of instability might not be accurate for certain non-repeatable features. For example, it is by no means clear as to whether vortex breakdown is a *bona fide* instability (*e.g.*, Gallaire *et al.* 2006), whether it is the result of a bifurcation process (*e.g.*, Gelfgat, 1996), or whether it is a combination of both (Leclaire & Sipp, 2010).

*Are they really vortices?*

Although when one naively observes the flow, the rings seem clearly to be vortex rings and although we were able to associate a circulation to at least the primary vortex, nonetheless the reasons offered for claiming that the sighted rings are indeed vortices may seem somewhat tenuous. The vortices which Archer et al (2010) studied had Reynolds numbers of some 7000, substantially larger than that the 4000 of our impinging drop. Our ring Reynolds number derived from the measured circulation as Re = $\Gamma/\nu$ =100/1= 100, a good order of magnitude below. However, their rings were generated from beneath the surface whereas ours were generated from above by a mechanism which still needs to be clarified. Furthermore for this low Reynolds number the short-wave instabilities (e.g. the Tsai–Widnall–Moore–Saffman instability) would not be expected. Although the long-wave instabilities (e.g., the Crow instability) may be related to the observed waves, more information, such a wavelength, growth rate etc. would be required to support the observed rings indeed to be vortex rings. It is thus direct observation coupled with the measured circulation which allow us to speak of these as being bona fide ring vortices.

**Conclusion**

When a drop falls into a shallow pool whose depth is of the order of the drop-radius two pairs of rings are generated. Two inner ring vortices of opposite sign are generated at the edge of the collapsing crater region, while two outer ring vortices (also of opposite sign) are 'laid off' by the expanding wave. In the inner pair the outer vortex predominates while in the outer group it is the inner vortex, which persists. The inner vortex first appears some three drop radii from the centre, the outer vortex at some seven drop radii from the centre (the question of scaling is not clarified). Although the inner vortex appears slightly before the wave vortex, the precise moment of their births is uncertain. With time the inner vortex shrinks slightly while the outer vortex grows. They both effectively cease to turn after some 20 msec. The characteristic turning of this stationary decaying ring can be followed with the naked eye even after the wave has passed.
The abovementioned vortices are most conspicuous for pool depths between 0.6 mm and 1.7 mm. Although we made an estimate of the circulation of this ring, since it grows and decays we cannot say whether this is the maximum circulation during its lifetime and it would be interesting to carry out a systematic investigation into its temporal development. A further direction for future study is the investigation of the dynamics of the dendrite, the ring vortex instabilities as well as mixing dynamics: Although the rings are so short-lived their mixing effect lingers like a brush-stroke on canvas, with its corresponding so aesthetically pleasing effect.

**Literature**

Archer, P.J., Thomas, T.G., Coleman, G.N. 2010 The instability of a vortex ring impinging on a free surface. *J. Fluid Mech.* **642**, 79-94